# The Emergence and Dynamical Evolution of Complex Transport Networks from Simple Low-Level Behaviours

Jeff Jones

Centre for Unconventional Computing,
University of the West of England, Bristol, BS16 1QY, UK

jeff.jones@uwe.ac.uk

The true slime mould *Physarum polycephalum* is a recent well studied example of how complex transport networks emerge from simple auto-catalytic and self-organising local interactions, adapting structure and function against changing environmental conditions and external perturbation. *Physarum* networks also exhibit computationally desirable measures of transport efficiency in terms of overall path length, minimal connectivity and network resilience. Although significant progress has been made in mathematically modelling the behaviour of *Physarum* networks (and other biological transport networks) based on observed features in experimental settings, their initial emergence - and in particular their long-term persistence and evolution - is still poorly understood. We present a low-level, bottom-up, approach to the modelling of emergent transport networks. A population of simple particle-like agents coupled with paracrine chemotaxis behaviours in a dissipative environment results in the spontaneous emergence of persistent, complex structures. Second order emergent behaviours, in the form of network surface minimisation, are also observed contributing to the long term evolution and dynamics of the networks. The framework is extended to allow data presentation and the population is used to perform a direct (spatial) approximation of network minimisation problems. Three methods are employed, loosely relating to behaviours of *Physarum* under different environmental conditions. Finally, the low-level approach is summarised with a view to further research.

## Introduction

The true slime mould *Physarum polycephalum* naturally prefers shaded or dark areas for its habitat. However, since the work of Nakagaki et al. in 2000 [19] on the maze solving abilities of the organism, the apparently simple unicellular multi-nucleate mass has been in the research spotlight as an unconventional model of computation and complex network organisation. The organism is able to form highly complex, adaptive and fault tolerant networks based upon only very limited local and distributed sensory and movement capabilities. Such behaviours are highly desirable in an era where computation is becoming ever more distributed and networked. The mechanisms *Physarum* utilises to naturally perform these computations were discussed by Nakagaki in [18] which were based on experimental observations (for example greater flow through shorter, thicker networks, quicker shrinkage of longer tubes vs. smaller tubes, the closure in open-ended tubes or 'dead ends'). These observed behaviours combined with assumptions regarding transport flux and hydrostatic pressure for directed flow in tubular networks [16] led to a mathematical model [22] which was used to perform shortest path computation for road network navigation. The network dynamics of *Physarum* in response to multiple food sources was investigated in [17] and the response of the organism tended to favour networks that achieved a balance between total length and network fault tolerance, as stated in terms of recent research into network stability and complexity by Strogatz [21]. The direct use of *Physarum* to construct spanning trees was performed by Adamatzky in [2] and the behaviour of *Physarum* was compared to a membrane-bound reaction-diffusion transport in [3]. In the same paper the authors noted that the availability of food resources was critical for the development of travelling localisations (a foraging-like behaviour by plasmodium extension emerged when food resources were low, compared to a uniform circular expansive growth when food resources were high).

*Physarum* is certainly not alone in the construction of complex patterned networks. Examples abound for organisms which differ greatly in the size, scale and individual motility at which the networks emerge (for example patterns in bacterial communication networks [6], plant leaf stomata dynamics [20], fungal mycelium growth [13], dynamic communication networks in insects [7] and vasculogenesis in human blood transport networks [5]). Although the individual organisms use very different physical methods to construct and re-model their respective networks, can the self-organisation processes that drive the original network formation and subsequent evolution be derived and synthesised?

This paper presents a synthetic, *bottom-up*, approach to the study of the emergence of *Physarum*-like transport networks, where the macroscopic network construction and evolution arise as emergent phenomena from low-level microscopic interactions (Figure 1). Using this approach, no underlying assumptions are made regarding the substrate or physical processes underlying the networks and only very simplistic (but, hopefully, plausible) local communication mechanisms (chemotaxis and diffusion of information) are used in an attempt to demonstrate the potential generality of the approach.

One disadvantage of creating a network model based on emergent behaviour is that the emergence and evolution of the network is very sensitive to changes in the individual particle behaviour and the effect of small changes to low-level behaviours on the global network behaviour is not always predictable. However, bottom-up models are based on fewer assumptions about the physical processes underlying the network formation which may improve their generality to other instances of complex network formation compared with top-down models.

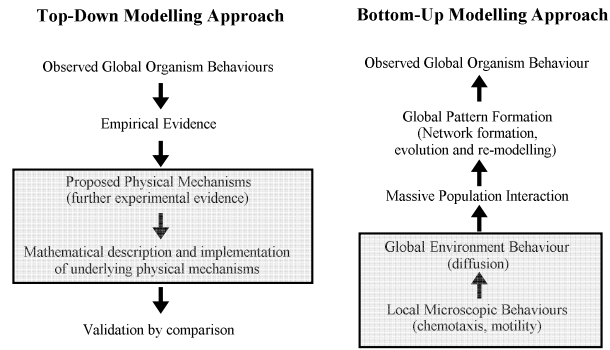

**Figure 1: Top-down vs. Bottom-up approaches to modelling of complex systems.**

(Shaded area represents positions where models can be adjusted)

Previous research into low-level mechanisms for network problems was stimulated by Dorigo's use of ant colony inspired behaviours for the Travelling Salesman Problem (see [10] for a summary) and the initial application has been extended to include efficient packet transmission in network routing [9], the Steiner tree problem [8] and efficient selection of pre-computed traffic routes [23]. Although inspired by emergent behaviour in ant colonies, and adhering to the tenets of emergent computation (locality, simplicity and population level emergent behaviours) most of the previous research, unlike real ant behaviour, does not actually adopt a directly spatial approach to the solution of the problems. The problems are typically represented by weighted graphs and the weighting of the graphs is used to probabilistically influence the selection of edges/routes (which are subject to reinforcement by 'pheromone' deposition). Direct spatial implementations have been used, however, local random walks for constructing minimum spanning trees were explored in [1] and reaction-diffusion mechanisms for network routing and load balancing were considered by Adamatzky and Holland in [4].

The direct spatial representation of networking problems presents challenges in terms of memory storage (since empty space must be represented between the nodes) and spatial resolution (a discrete spatial representation will be limited by its resolution) when compared to graph representations. However, the direct representation presents the opportunity for emergent mass behaviours to show their influence. For example, how does overcrowding or the sudden availability of space affect the mass transport? Such possibilities can only be explored with a spatial representation and surprising global mass transport behaviours can emerge from the confinement interactions, such as in the particle model of Heigeas et al. [12].

For this research a multi-agent framework was used to generate the phenomena using simple particle-like behaviours and simple diffusion processes. The framework consists of a *landscape* consisting of a rectangular two dimensional array of height-field values (for example a digitised greyscale image) that is presented as a habitat for a population of simple, mobile reflexive software agents. The data that comprises the landscape drives the behaviour of the individual agents and the 'output' of the framework is the historical record of the distortion of population density which is recorded in data structures isomorphic to the original data landscape. The historical behaviour of the population represents a *collective perception* of the original landscape data and the paradigm and framework has been demonstrated in exemplar applications for image processing (see [15] for a more complete exploration of the data driven approach and framework parameters), and as a method for reconstructing absolute achromatic brightness in the human visual system [14]. To study the emergence and evolution of transport networks, a simplified version of the framework was used with zero, or at most very limited, stimulation from the landscape habitat and an emphasis on chemotactic sensory behaviour and diffusion.

The remainder of this report contains a description of agent morphology, agent behaviour and the diffusive environment behaviour in section 2. The spontaneous formation of transport networks is illustrated in section 3 and the mechanisms underlying their formation and long term evolution are explained. Section 4 explores their application to network minimisation problems by the addition of very limited landscape stimuli to attempt to constrain the behaviour of the emergent networks. Three different methods are discussed, each loosely based on different observed 'modes' of *Physarum* behaviour (in terms of food availability, foraging ability and plasmodium size). Section 4 concludes with a discussion about networks with cycles and the possible mechanisms and advantages of cyclic network formation. The report concludes with a discussion of how the processes of spontaneous emergent network formation may be related to self-organising networks seen in physical and living systems. Due to the dynamical nature of network formation and evolution which cannot be easily represented in static image form, the reader is encouraged to refer to the supplementary video recordings of network formation and evolution related to the figure captions (http://uncomp.uwe.ac.uk/jeff/physarum.htm).

**2 Multi-Agent Framework - Agent Morphology and Behaviour.**

The agents follow simple stimulus-response behaviour, almost particle like in their simplicity. The general morphology of an agent and its underlying algorithm is illustrated in Figure 2. An agent occupies a single location in the environment, corresponding to a single pixel of a digitised image. Each agent is initialised at a randomly chosen unoccupied location and with a random orientation (from zero to 360 degrees, freeing the agent from the restrictive architecture of the underlying orthogonal image). The agent receives sensory stimuli from its environment (the trail map data structure) via three forward sensors and the agent responds to differences in the local environment trail levels by altering its orientation angle by rotating left or right about its current position. The agent is forward biased, i.e. only the sensors directly in front of the agent's current position are used to influence its behaviour, thus ensuring a continuous dynamic.

At each execution step of the scheduler every agent attempts to move forwards one step in the current direction. If the movement is successful (if the next site is not occupied) the agent moves to the new site and deposits a constant trail value. If the movement is not successful the agent remains in its current position and no trail is deposited and a new orientation is randomly selected. Note that the agent both deposits *and* senses the trail map, resulting in an autocatalytic paracrine mode of stimulus/response. The parameters for population size (%p), sensor angle from forward position (SA), sensor width (SW), sensor offset distance (SO), agent step size (SS), trail deposition rate (depT), and the agent's angle of rotation (RA) can be adjusted, generating different emergent network behaviours. The trail data structure is subject to a simple diffusion operator after every system step (one system step defined as the sensory and movement steps of every member of the population). The diffusion operator takes the form of a pseudo-parallel simple mean filter in a 3x3 kernel which is subject to an adjustable damping value to affect trail persistence.

```
[Motor stage]
- Attempt move forwards in current direction
- If (moved forwards successfully)
        Deposit trail in new location
- Else
        Choose random new orientation
[Sensory stage]
- Sample trail map values
- if (F > FL) && (F > FR)
        - Stay facing same direction
        - Return
- Else if (F < FL) && (F < FR)
        Rotate randomly left or right by RA
- Else if (FL < FR)
        Rotate right by RA
- Else if (FR < FL)
        Rotate left by RA
- Else
        Continue facing same direction
```

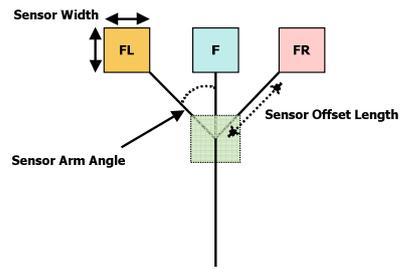

**Figure 2: Agent low-level behaviour pseudocode and sensory morphology**

Individual agents (without sensory stimuli) behave in a pseudo random walk whose stochastic influences include the initial random start location, random initial orientation and the random selection of new orientation in the event of an unsuccessful movement. The addition of sensory paracrine stimuli (via agent trail deposition) transforms the population behaviour into a guided random walk which can be further influenced by the presence of landscape stimuli.

## 3 Spontaneous Emergence of Dynamic Networks

The experiments presented in this document were based around the following parameter settings for the framework shown in table 1. When a parameter is changed to illustrate different network behaviours it is indicated in the text.

| Parameter | Value | Description |
|---|---|---|
| Image | 200 x 200 pixels | Blank image |
| %p | 5 | Population as percentage of image area |
| SA | 15 degrees | FL and FR Sensor angle from forward position |
| SO | 15 pixels | Sensor offset distance |
| RA | 45 degrees | Agent rotation angle |
| SW | 1 pixel | Sensor width |
| Boundary | Periodic/Fixed | Diffusion/Agent boundary conditions |
| depT | 5 | Trail deposition rate (constant) |
| damp | 0.1 | Diffusion damping factor |
| Node weight | Typically 0.01 to 0.1 | The weight applied to the input of node sources to the trail map (section 4) for network problems |

**Table 1: Base parameter values for framework experiments**

The effect of the base parameters is the spontaneous emergence of a regular network as shown in Figure 3.

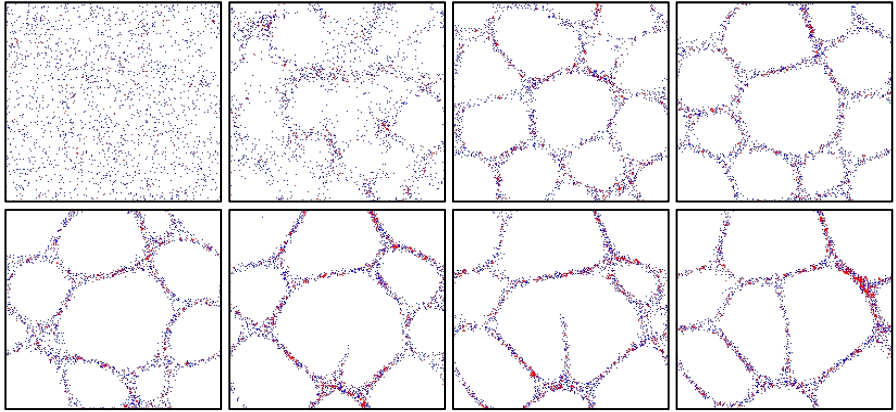

**Figure 3: Emergence of a persistent dynamical network**

(1-4) Top row, left to right
(5-8) Bottom row, left to right

Agent particles are shown as small specks. The initially random distribution and movement (Figure 3, 1) appears to undergo a phase transition when the chemotactic behaviour of the agents forms trails (2). The paths condense into a regular network (3) which itself undergoes deformations such as the closing of smaller circular structures (3-5). Also of note is the appearance of branching structures which appear by their movement to forage forwards by streaming and ultimately joins together two distant parts of the network (6-8). The network evolution with the base parameter settings never settles into equilibrium and the network topology continues to change although the number of agent particles representing the network remains constant. The video recordings illustrate that movement along the network is bi-directional.

Increasing the sensor angle to 45 degrees generates a network that is able to settle into an equilibrium state (closing of circular structures but without the branching streams) as shown in Figure 4 which illustrates the trail intensity left by the agent movement. Note that trail intensity closely corresponds to agent position and the two are used interchangeably in the remaining figures.

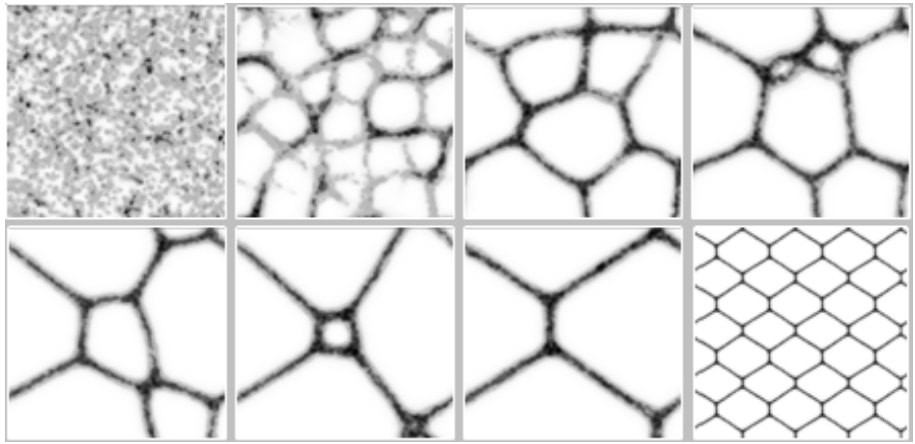

**Figure 4: Emergence of a dynamic equilibrium state with sensor angle of 45 degrees.**

(1-4) Top row, left to right
(5-8) Bottom row, left to right

The stable state[1] (image 7) can be tiled to reveal the effect of the periodic boundary conditions and shows a roughly hexagonal honeycomb-like pattern (final column, bottom row). The properties of a hexagonal honeycomb lattice have been shown to be a minimal surface connector [11] in two dimensions and the evolution of the emergent network appears to be a minimisation process. The angles of the honeycomb structure were found to be 120 degrees but had unequal path lengths (117 pixels vs. 43 pixels) in the ratio of 2.72 to 1 and the causes for the unequal path lengths that emerge have not yet been found. Although the formation of the honeycomb structure was affected by extremes of parameter variation (for example very small sensor offset distances resulted in only local networks), when global networks did emerge the 120 degree angle was surprisingly not shown to be affected by differences in sensor angle (SA) or agent rotation angle (RA).

---

[1] Although the pattern appears to be static, the apparent stability is in fact the continuation of the dynamical mass transport by the agent population.

If, when in the stable state, the sensor angle was reduced to 15 degrees, the foraging streaming behaviour re-emerged and the stable state evolved into the previously seen dynamical patterns. Resetting the sensor angle to 45 degrees provoked a return to the stable regime. During the return to stability any sprouting branches gradually shrunk back to be absorbed within the main mass of the network (see supplementary recordings). With a sensor offset distance (SO) of 15 pixels, the threshold value for the emergence of the streaming behaviour was a SA of 15 degrees. At angles greater than this, the stable regime returned.

To further explore the surface minimisation behaviour the evolution was performed without periodic boundary conditions for the agent movement and diffusion and the structures shown in Figure 5 emerged. The evolution of the network showed minimisation towards the centre of the area but there was adhesion of the network to the four corners. Further investigations found that the adhesion was not an inevitable part of the network formation but was due to imbalanced inputs to the agent's sensors when facing the corner of the screen: When facing in this orientation the agent received no trail input from the left or right sensors since these were outside the border area, but the front sensor still received a trail input as this sensor was still within the border area.

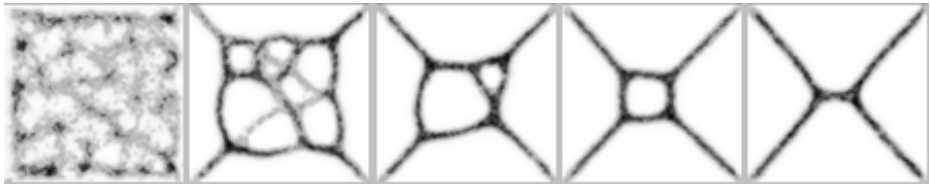

**Figure 5: Surface minimisation on fixed boundaries with adhesion in the corner areas.**

To remove the adhesion artifacts, the agent algorithm was adjusted so that if *any* sensor was outside the image area, then the agent would arbitrarily turn right in an attempt to avoid becoming confined at the corner areas. The result of removing the adhesions can be seen in Figure 6 (which repeated the same experiment as in Figure 5 with the addition of adhesion removal) and illustrates the minimisation of the network into its minimal surface area, a circle.

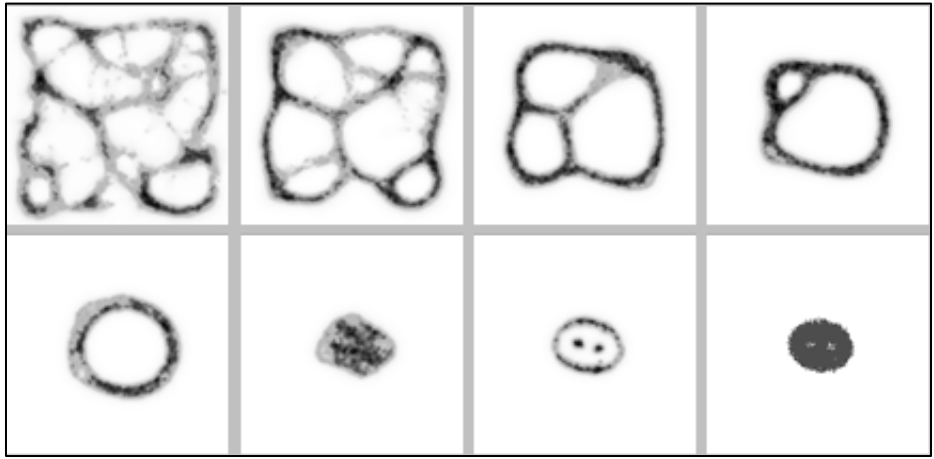

**Figure 6: Surface minimisation without adhesion**

(1-4) Top row, left to right
(5-8) Bottom row, left to right

The evolution of the minimal surface network begins with the closing of the smaller circles (2-4) and the shrinkage of the remaining single circle (5-7) by what appear to be tension effects. The trail representation of the minimal surface shows a ring with two smaller particles within (7). In fact the agent distribution (8) of the same stage shows an almost solid mass of agents. The ring in the trail pattern occurs because only those agents that are free to move (those at the periphery) will deposit trail. The bright points of trail within the circle are caused by the momentary opening and closing of gaps within the solid mass where movement can temporarily occur.

### 3.1 Mechanism of Transport Network Formation

The emergence of the dynamical transport networks and apparent surface tension effects arise from the interaction between agent movement, the positive chemotaxis response of the individual agents (movement towards a concentration gradient and local paracrine deposition and sensory function), the diffusion of the trail 'chemical' and the morphology of the agent sensory structures.

The agent morphology (Figure 2) is itself similar to a branching structure and the behaviours in the pseudocode either generate continuing forward movement or left/right branching movement. The initial aggregation phase is caused by the forward directed movement of the agents and the diffusion of trail as the agents move. The forward movement leaves a conical wave of diffusing trail behind the agent (Figure 7, left). If another agent is behind this trail wave, its sensors will pick up the trail and the agent will fall behind and track the movement of the first (Figure 7, right).

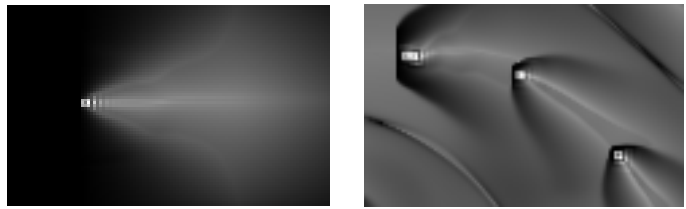

**Figure 7: Initial aggregation caused by diffusion and chemotactic movement**

The aggregation continues as greater numbers of agents become drawn in to the emerging paths due to the greater amount of trail being deposited by the train of agents. Networks form when different trains of agents cross paths. If no diffusion of trail occurs, networks of trails will still form but no network minimisation or reconfiguration occurs.

When the trails are subject to a diffusion pressure the minimisation behaviour of the network occurs when all agents are contained within paths. As agents only deposit trail after a successful forward movement, the agents at the centre of a network path are more likely to be blocked and leave no trail, whereas agents at the boundary region are more likely to be free to move (Figure 8, left and middle). This difference in trail deposition creates areas of instability in the network paths. The closing of the circular structures occurs because agents on the inside of the circle have a shorter distance to traverse than agents on the outside of a circle. The greater flow caused by the shorter distance results in more trail being deposited at the inside of the circle and ultimately more attraction of the agents towards the shorter path (since agents are attracted to the strongest *amount* of diffusing trail chemical). Also, the inside of the circle is an unimpeded 'track', whereas the outside of the circle (Figure 8, right) is interrupted by 6 spoke-like feeding paths. This also contributes to the greater flow of agents around the inner circle boundary. The flow is bi-directional and the greater flow causes greater agent movement and minimisation of the circle in a positive feedback manner (smaller circle > greater flow on inner circle > smaller circle etc.) until the circle is entirely closed.

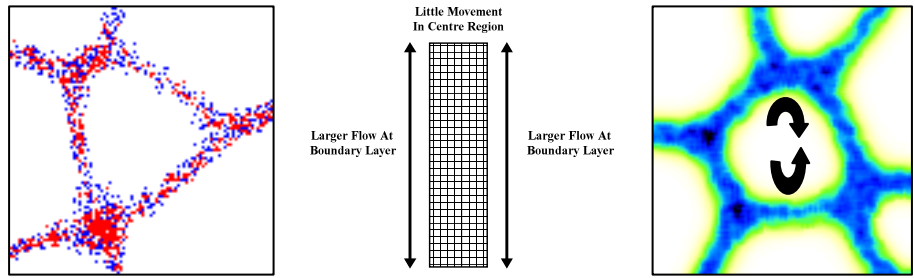

**Figure 8: Differences in agent flow at the inside and outside of a network path cause circular loop contraction.**

Left:       Network 'walls' actually consist of multiple layers of agents.
Middle:     The internal layers are more restricted but outer layers exhibit boundary flow
Right:      Circular contraction of network caused by greater internal boundary flow.

The stability of the network (with a 45 degree sensor angle) is ensured by the fact that the number of agent particles remains constant. Therefore any increase in flow of agents in one part of the network must result in a decrease in agent flow in another part of the network. The result is a network that minimises the gross changes in agent distribution whilst maximising agent flow since agent flow is greatest with the minimum distance. This ultimately results in the hexagonal structure.

The switch to foraging and streaming behaviours, and the dynamical network that ensues when the sensor angle is set at 15 degrees, is explained by the fact that the agent rotation can now remove the (narrow) sensors from the current flow of trail that the agent was following. The initial stages of network formation are identical to the stable 45 degree condition, with path formation and closing of circular structures. When the network achieves a degree of stability the branching and streaming behaviours are observed. The branching structures emerge from the side of a network path boundary flow, presumably due to instabilities caused by agent-agent collisions (Figure 9, 1 and 2). If the random direction selected is at approximately 90 degrees to the flow (i.e. out of the range of the strong path signal) then a small stream of agents will start to 'explore' by forward movement the empty space (3-5). The streaming is bi-directional, surging forwards, then backwards. When the gap between network paths is almost crossed, the sideways flow disrupts the boundary flow of agents in the opposite side of the gap and (remembering that agent's sensors are some way from the actual agent 'body' position) agents from the opposite side will start to branch outwards to complete the path (6-8). When the gap is crossed, a circular structure results which is closed in the manner described previously, unless a further instability emerges.

Further instabilities in the boundary flow are very likely because the sudden surge of flow into one part of the network results in a decrease in flow in another part of the network. The end result is a highly dynamical network whose topology and structure constantly changes due to movement of its component parts.

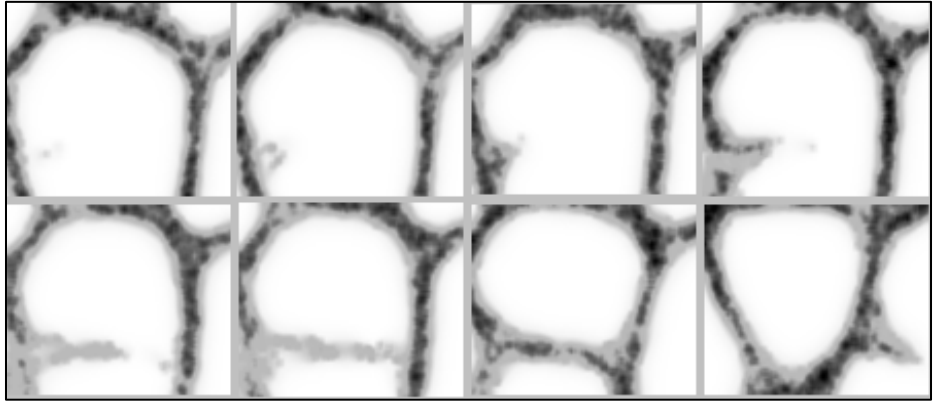

**Figure 9: Evolution of branching and streaming behaviour when SA is 15 degrees.**

(1-4) Top row, left to right
(5-8) Bottom row, left to right

## 4 Constraining the Dynamic Networks for Useful Behaviours

The emergence of transport networks from the low-level particle activity appears to possess features in common with living systems (foraging, streaming and minimisation in *Physarum*) and physical systems (minimisation by surface tension). Attempts were made to harness the emergent behaviour for useful purposes.

One of the central difficulties in unconventional methods of computation is how to provide a representation of the problem inputs and how to read the problem outputs. In their experiments on using *Physarum* for network problems, Nakagaki et al. used food sources as the input nodes. The pre-starved plasmodium then filled the entire problem area before minimising its transport structures (internal veins in the plasmodium) to maximise contact with the food sources and to minimise overall network length whilst providing a fault tolerant network [18].

Presenting problem inputs to the multi-agent population was achieved by representing the problem nodes as constant diffusion sources within the trail field. The input nodes provided chemotactic attraction to the agent particles. The desired aim was to utilise the foraging and minimisation behaviour to explore network optimisation and thus the problem 'output' would be the final state, or persistent stable dynamic state, of the emergent network. Since the agent population does not actually 'consume' the sources of diffusion, three different strategies were used to explore the problem solving abilities of the emergent networks:

1. The initialisation of the entire agent population of the landscape (filamentous shrinkage).
2. The gradual initialisation of the population only at the problem node points (filamentous foraging).
3. The initialisation of a massive population on the entire landscape and the gradual reduction of population size (plasmodial shrinkage).

Simple problem datasets were initially used, similar to those in Nakagaki et al. [18] and examples of the problem sets are shown in Figure 10.

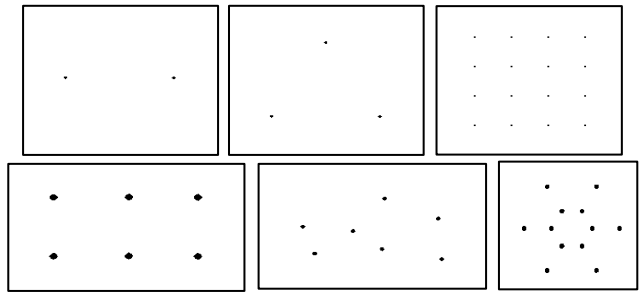

**Figure 10: Network length problem datasets**

Dots represent parts of landscape where diffusive trail is introduced at the network nodes.

### 4.1 Filamentous Shrinkage Method

The filamentous shrinkage approach initialises a small population of agent particles onto the landscape. The landscape is also stimulated by the diffusion sources indicating the problem nodes. The agents begin to form network paths resembling filaments and at this stage the formation of the network occurs as in previous experiments but the surface minimisation is constrained when the population is attracted to node points and the network becomes 'snagged' on the nodes. Minimisation of the network (such as the closing of circular structures) can still occur as shown in Figure 11 but the network remains attached to the node points. The network spanning the nodes in the following examples appear to approximate Steiner minimum trees, i.e. the shortest overall connectivity between the nodes with the possible addition of extra Steiner nodes to reduce overall network length.

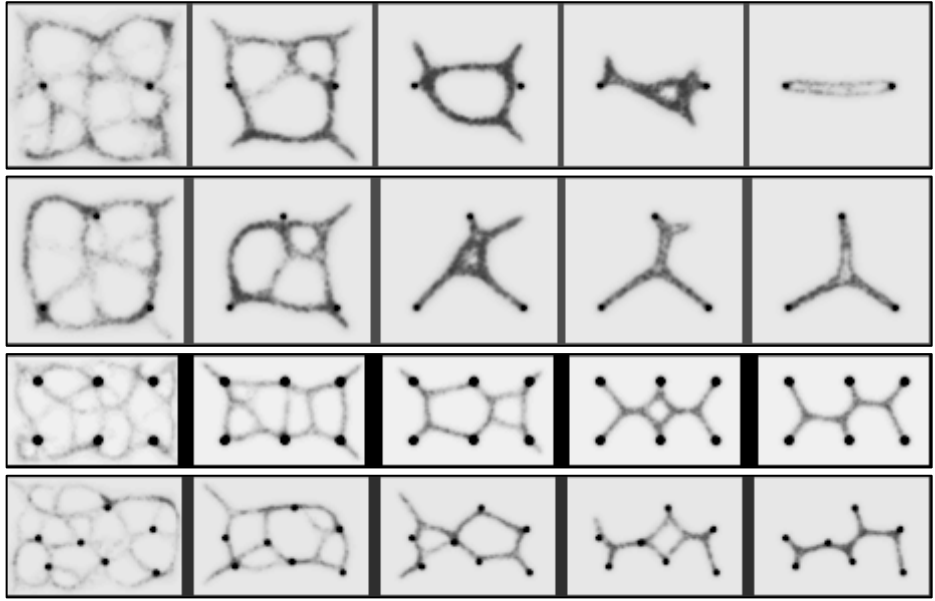

**Figure 11: Network minimisation by filamentous shrinkage on 2, 3, 6 and 7 node networks.**

Different problem on each row. Network evolution from left to right.

The node sources of diffusion are copied to the trail map at every system step and diffuse in the same manner as the particle trails. The strength of the node points can be subject to a weighting value that affects the attraction of the agents to the nodes. The filamentous shrinkage method is affected by the diffusion damping, the trail deposition level and the weighting value applied to the diffusion node points. If the attraction of agents to the node sources is too strong then the loop structures will not be minimised. If the attraction to the node points is too weak the network will 'tear' and lose contact with one or more nodes. If the sensor angle is reduced to 15 degrees the network will not remain at the minimal configuration but will search for local variants of the minimal configuration forming complex dynamical patterns. Similar foraging behaviour can be induced by a diffusion damping that is too strong. The effect of a very strong damping of diffusion is akin to the searching propagation of *Physarum* under poor nutrient conditions. Examples of such erroneous or incomplete minimisation are illustrated in Figure 12.

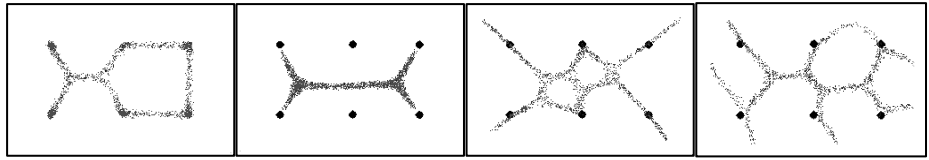

**Figure 12: Incorrect minimisation of network using filamentous shrinkage.**

Left to Right:
  Too strong attraction to node points – loops are not closed because of strong node attraction
  Too *weak* attraction to node points – network is 'torn' from nodes
  Too narrow sensor angle generates foraging dynamic network
  Too strong diffusion damping also generates foraging dynamic network

The importance of the diffusing node sources is illustrated in Figure 13 which shows the consequences of removing the node sources. Without these sources of attraction for the agents, the tree collapses as the minimal surface is found.

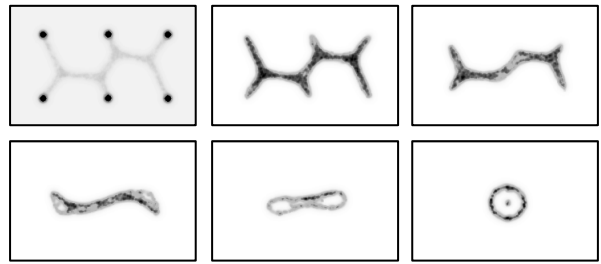

**Figure 13: Network collapse when network node sources are removed**

The relatively low population sizes used for filamentous shrinkage ensure that the initial random placement and direction of agent particles can affect the minimisation of the network. The minimal network was typically found regardless of the initial 'shape' of the emergent filament paths. However, the stochastic factors affecting the initial network formation (before the minimisation of the network between the nodes) ensures that the correct minimisation of the network cannot always be guaranteed, especially in networks with a large number of nodes.

### 4.2 Filamentous Foraging Method

As with the filamentous shrinkage method a relatively small population sized is used. The foraging method initialises the agents with random orientations and positions but the initialisation positions are restricted to the network node source locations. Since there is not enough room on the node points for the entire population, the population cannot be initialised at once and a gradual introduction to the landscape is necessary. As the agents are introduced onto the landscape they stream out from their node source making connections to nearby nodes generating filament-like networks.

Due to the strong attraction of the network nodes, there is the possibility that agents may remain constrained within the area of the node points rather than foraging outwards. This was avoided by temporarily decreasing the sensor angle to 15 degrees to increase the foraging behaviour and then resetting the angle to 45 degrees to stabilise the dynamic network. An alternate possibility would be to reduce the weighting value of the node diffusion sources, or to increase the step size distance (SS – or how far the agent moves at a single 'stride').

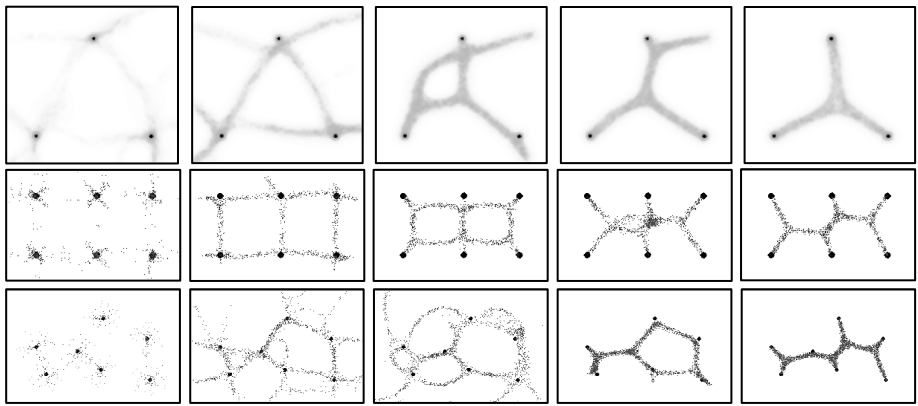

**Figure 14: Network minimisation using the filamentous foraging approach**

Note that in the early stages of the network formation using the foraging approach, the Delaunay triangulation (DTN) is formed (Figure 14, second column) which is especially noticeable when the images consisting of regular shapes (triangle, domino shape). This is because the nodes in the regular shapes are the same distance from their neighbours and thus equally attractive. Once the DTN is formed the network continues to minimise to an approximation of the Steiner tree.

### 4.3 Plasmodial Shrinkage Method

The plasmodial shrinkage method was an attempt to overcome the limitations that small population sizes have on network minimisation. The method is similar to the behaviour of *Physarum polycephalum* where the pre-starved organism grows to fill the entire problem area and then shrinks in size to maintain optimal connections between the food sources. The idea behind this methodology for the emergent multi-agent networks is that the stochastic influences on the initial network formation that hampered small population sizes would ameliorated by the wide coverage of a large population.

A very large population size (>40% of the image area) is required for the plasmodial shrinkage method. The population is initialised on the entire landscape and a transport network starts to emerge. This network takes the form of a large sheet with holes (Figure 15, 1$^{st}$ column). The holes close up via the previously described mechanism of minimal surface flow shrinkage. The sheet now encompasses the entire set of nodes (2$^{nd}$ column).

When the trail pattern becomes a stable, amorphous, uniform sheet, agents are removed from the population at random (p0.001 which provides enough time for the agents to adapt to the decrease in population size – a greater frequency of removal presents the risk of re-occurrence of holes in the sheet). In this method when an agent has been removed it can only be re-initialised on a part of the image that contains a network source node. Since the nodes are very small in area, the effect is to reduce the population in size. The population responds to the gradual fall in population size by shrinking the surface area of the amorphous sheet. The attraction of the network node sources ensures that the network flow remains connected to these sources and the shape of the sheet becomes affected by the structure of the node points. The final structure (at a population size similar to the previous methods ~3-5% of image area) reveals the minimal network (5$^{th}$ column).

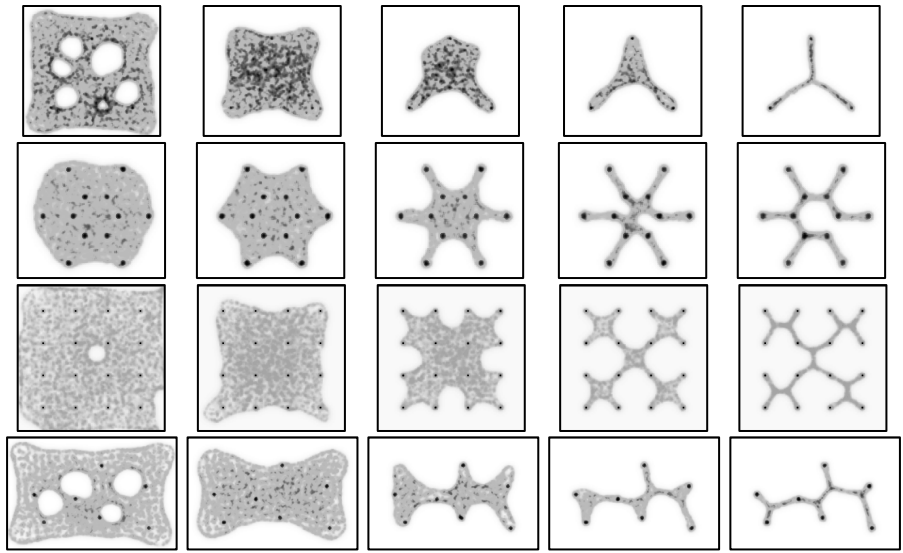

**Figure 15: Evolution of the plasmodial shrinkage method.**

(Dark spots indicate node diffusion points. Small flecks indicate dynamic fluctuations in the density of the trail 'sheet'. See video recordings for illustration)

If the reduction of population size is not as drastic, the sheet will deform to encompass all of the points of the network and illustrate the concave hull of the set of points. Also, for relatively simple networks (those without nested internal regions) tracking from node to node by following the outside of the sheet will reveal the shortest tour of the Travelling Salesman Problem (Figure 16).

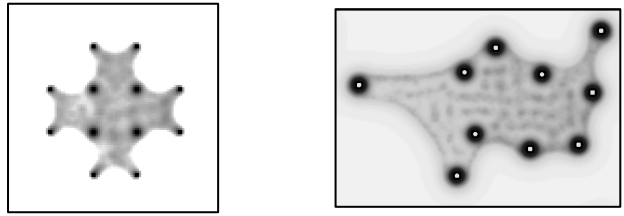

**Figure 16: Approximation of concave hull and TSP tour using plasmodial shrinkage.**

### 4.4 Cyclic Networks and Networks with Continuous Remodelling

The results of Nakagaki et al. on *Physarum* networks with numbers of food sources greater than or equal to three showed a response that combined short overall network length with resilience to accidental path disconnection – a fault tolerant network [17]. Typically the shape of these network patterns was based on a modification of Steiner minimum trees with the addition of one or more cyclic connections. The extra apparent 'cost' of cyclic connections in terms of network length appears to be offset by the increased reliability of the network to random disconnections. Nakagaki also reported variability in the specific type of network construct, grouping into three main types: Steiner type tree (SMT), Cyclic network (CYC) and combined variants (COM).

A similar behaviour was seen in the multi-agent emergent networks. On networks with a large number of nodes, instabilities caused by stochastic factors influencing the initial network paths (especially in the filamentous methods, but also occurring in the shrinkage methods) often resulted in non-optimal networks (in terms of path lengths) or networks with random disconnections (Figure 17, top row). However the cycles formed in such non-minimal networks would appear to enhance connectivity against the threat of accidental disconnection.[2] A continual remodelling of the network connections was apparent when the foraging-like behaviour afforded by reducing the sensor angle to 15 degrees was used. Figure 17, (bottom row) shows snapshot images of the dynamic remodelling behaviour seen in networks with large numbers of nodes and the foraging agent behaviour. The pattern of the network at the 'core' remains relatively stable and the outer connections are continuously remodelled. The emergent networks contain cycles and the foraging behaviour also prevented the random disconnection of network nodes.

---

[2] No quantitative analysis of this behaviour has yet been carried out due to difficulties in when deciding when to 'halt' the continuous dynamics within cyclic networks. Because networks with large numbers of nodes do not settle into a minimal state, the cycles undergo a continual process of disconnection and reconnection.

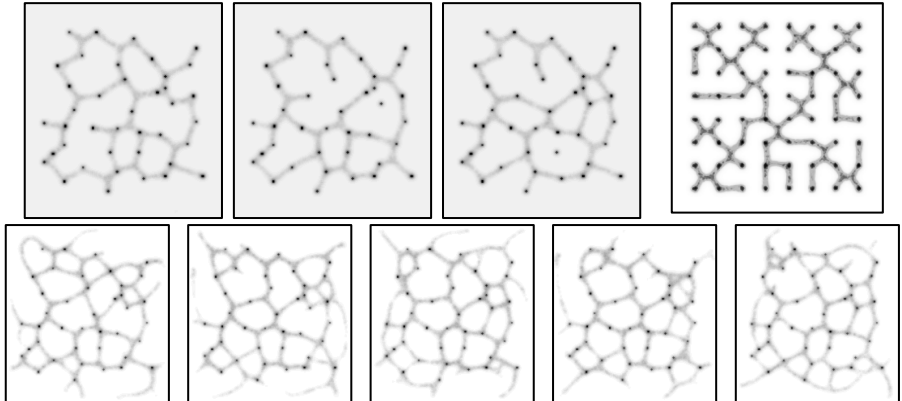

**Figure 17: Cyclic and erroneous minimisation and dynamic remodelling**

Top Row:
    Erroneous network minimisation using filamentous shrinkage with cycles formed (columns 1-3). Erroneous minimisation of 64 regular nodes using plasmodial shrinkage caused by instabilities in the deforming sheet (Column 4)
Bottom Row:
    Continuous dynamic network remodelling by foraging behaviour in filamentous shrinkage. Each image is taken every 1000 steps of the multi-agent scheduler.

## 5 Summary and Scope for Further Research.

A low-level, bottom-up approach to the modelling of complex transport networks was presented. The spontaneous formation of mass transport networks was observed from the spatial interactions of microscopic, locally specified, chemotactic (paracrine) sensory and movement behaviours, and a (globally applied to all sites) local mechanism of diffusion. Second-order emergent phenomena were observed (closing of circular structures and branching by streaming) that served to generate complex network evolution combined with a physical analogue of minimisation by emergent surface tension. The evolution of the network could be switched between the dynamic foraging behaviour to network minimisation by the adjustment of the sensor angle of the agents.

Network nodes were represented by diffusion point sources that attracted the individual agents and thus affect the network formation and minimisation. The evolution of the network was used to spatially compute approximations of minimum Steiner trees. Three methods were used, two of which (filamentous shrinkage and filamentous foraging) approximated chemotactic foraging behaviour in *Physarum* in nutrient poor conditions and a third (plasmodial shrinkage) which approximated the shrinkage of the (post-fed) plasmodium sheet to minimise surface area.

Non-minimal networks with cycles were formed with more complex data sets, suggesting that random perturbations (caused by uneven initial population distribution or random influences during the evolution of the system) affect the emergent network. By adopting the foraging behaviour via narrow sensor angle within complex data sets, the network length vs. fault tolerance trade-off that *Physarum* adopts, also seems to occur in the emergent transport networks (from initial observations). The complex networks undergo a continual process of dynamic remodelling. It is interesting to speculate if this evolved behaviour is perhaps due to the indirect *anticipation* of random path disconnection (i.e. random disconnections acting as a selection pressure to evolutionarily favour the foraging vs. stability compromise).

The low-level approach adopted in this paper attempts to provide generality to the study of transport networks. Rather than trying to create top-down approaches to how a particular organism (in this instance *Physarum*) performs the network behaviours, we concentrate on the actual emergence of the network structures themselves from very simple component parts. It is possible that variations in the sensory structure or simple particle algorithm behaviour may generate significant differences in the type of network behaviour that emerges, and in the long term dynamics of the networks. It is hoped that this approach may provide insights into the *common* processes under which network formation and evolution may occur in a wide range of complex biological transport networks.

## Bibliography.